\documentclass{optica-article}

\journal{opticajournal} 

\articletype{Research Article}

\usepackage{lineno}
\usepackage{wrapfig}
\usepackage{bm}
\usepackage{xcolor}
\usepackage{enumitem}
\usepackage{booktabs}
\usepackage{microtype}
\usepackage{multirow}
\usepackage{amsmath}
\usepackage{hyperref}

\newcommand{\luxcorerender}{{LuxCoreRender}}
\DeclareMathOperator{\mae}{MAE}

\begin{document}

\title{Double-Freeform Lens Design for Angular-Spatial Control of Light Fields}

\author{Yuou Sun,\authormark{1} Bailin Deng,\authormark{2} and Juyong Zhang\authormark{1,*}}

\address{\authormark{1}School of Mathematical Sciences, University of Science and Technology of China, Hefei, China\\
\authormark{2}School of Computer Science and Informatics, Cardiff University, Cardiff, UK}

\email{\authormark{*}juyong@ustc.edu.cn} 


\begin{abstract*} 
Precise simultaneous control of both angular and spatial light-field distributions remains a longstanding challenge in optical design, often requiring complex multi-element configurations.
In this work, we propose a compact single-lens solution that achieves unified angular-spatial modulation through the co-optimization of double freeform surfaces.
The problem is formulated as an extended caustic design that enforces prescribed irradiance patterns on two distinct receptive planes, where the dual-plane constraint implicitly defines the directional characteristics of the light field while preserving spatial accuracy.
This framework eliminates the need for auxiliary optical components while delivering performance comparable to that of conventional multi-lens systems.
Comprehensive numerical simulations verify the method’s effectiveness, demonstrating accurate and stable control of both angular and spatial light-field properties.
The proposed approach establishes a practical foundation for compact, high-performance optical systems and provides a promising route toward integrated angular-spatial light-field engineering.

\end{abstract*}

\section{Introduction}
Achieving simultaneous control over both the angular and spatial characteristics of light fields remains a central challenge in computational optics and illumination design. Angular control dictates how light propagates through space, whereas spatial control determines the irradiance distribution on a receptive plane. Although many studies report impressive results, they often address these two objectives separately, leaving joint control within a single compact element challenging to achieve in practice.

This separation becomes a practical limitation in applications requiring both “what pattern” and “in which direction”. In particular, most freeform-lens and caustic-design formulations prescribe a target irradiance on a single receptive plane only, leaving the outgoing angular distribution underconstrained and limiting flexibility for tasks such as controlled divergence, beam steering, or maintaining collimation. A common workaround is to stack multiple optical components (such as lens arrays~\cite{li2020beam}, microlens stacks~\cite{prossotowicz2021dynamic} or multi-plane light conversion systems~\cite{kupianskyi2023high}) or to cascade spatial light modulators~\cite{pinilla2022hybrid} and diffractive elements~\cite{soshnikov2023design}. While these multi-stage solutions address the missing degrees of freedom at the system level, they complicate alignment and calibration, increase optical path length, and raise both fabrication cost and system complexity.

Meanwhile, substantial progress has been made along each axis of control individually. 
On the angular side, recent methods can explicitly shape propagation behavior. For example, Wei et al.~\cite{wei2023sculpting} sculpt optical fields into prescribed three-dimensional caustic trajectories, enabling fine control of where and how light travels through space. 
On the spatial side, freeform surface design has been widely used to produce prescribed irradiance patterns on a receptive plane. A representative example is the physical-optics-aware freeform light shaping model of Yang et al.~\cite{yang2020light}, which realizes complex caustic images under uniform parallel illumination. 

Beyond optics, similar light-control goals—caustic design in particular—have also been widely studied in computer graphics~\cite{kiser2013architectural, yue2014poisson, schwartzburg2014high, meyron2018light}. These approaches often represent the freeform geometry as a triangle mesh and optimize the illumination pattern by directly updating vertex positions. Their formulations are typically mesh-based and ray-driven, closely related to the differentiable mesh-and-ray pipelines that we build upon in this work.

A natural question then arises: how can joint angular and spatial control be achieved without resorting to cascaded hardware? Double-freeform optical elements provide a promising direction because two surfaces offer additional degrees of freedom beyond single-plane irradiance control. Early work demonstrated beam shaping with double-freeform surfaces, e.g., transforming a Gaussian beam into a uniform distribution~\cite{feng2013beam}. Subsequent methods extended this idea to simultaneously realize a prescribed wavefront and receptive-plane irradiance~\cite{feng2013designing}, typically by first constructing a suitable ray mapping and then recovering the two freeform surfaces accordingly. Subsequent methods extended this line into Monge--Amp\`ere--based formulations, improving numerical stability and generality for laser beam shaping~\cite{zhang2014double}, where the double-freeform design problem is cast into a PDE-based framework coupling energy conservation, refraction, and optical-path constraints. The concept was later broadened to illumination design~\cite{bosel2018double}, with PDE formulations generalized to prescribed irradiances and wavefronts under more general input/output conditions. Complementary to these PDE-based approaches, supporting-quadric methods provide a more constructive route: instead of solving directly for the freeform surfaces through a global PDE, they build the solution geometrically from local supporting quadrics and have also been shown effective for designing double-surface refractive elements with prescribed irradiance distributions and wavefronts~\cite{bykov2021supporting}. Another related line is iterative wavefront tailoring (IWT), which introduces an intermediate outgoing wavefront and iteratively updates it to simplify freeform design for prescribed irradiance~\cite{feng2019iterative}. This idea was later extended to optical-field control, where the iterative framework is used to regulate not only output irradiance but also phase-related field behavior~\cite{feng2021iterative}. More recent differentiable frameworks further enhanced modeling flexibility~\cite{tang2024differentiable}, enabling gradient-based optimization of compact double-freeform lenses under more flexible forward models. 
Related two-target designs have also been explored in other optical settings. Braam et al.~\cite{braam2025inverse} considered the inverse design of two reflectors for a parallel-to-two-target system by deriving optical mappings to the two targets and reconstructing the reflector surfaces through generating functions. This framework was later extended in~\cite{Braam:26} to handle both refractive and reflective systems under a broader class of zero-\'{e}tendue sources, unifying the treatment of lens and reflector configurations within a single design methodology.
Importantly, Romijn et al.~\cite{romijn2021generating} showed that a double-freeform lens design for a prescribed far-field intensity admits a free parameter that produces a family of distinct solutions; this indicates that the two-surface geometry offers more flexibility than a single target constraint requires, which can be exploited to regulate directionality or to satisfy constraints on multiple planes.

Motivated by this insight, we formulate a design approach that exploits this additional flexibility to achieve simultaneous angular and spatial control via dual-plane constraints. Our goal is to design a single double-freeform lens whose refracted field forms prescribed irradiance distributions on two distinct receptive planes, $A$ and $B$. By doing so, we shift the emphasis from generating alternative solutions for one-plane shaping to leveraging the additional degrees of freedom for multi-plane constraints. Moreover, the resulting coupling between planes naturally links angular regulation to spatial shaping between $A$ and $B$. In contrast to methods that recover the double-freeform geometry indirectly through ray mappings, PDE solutions, supporting quadrics, or iteratively tailored intermediate wavefronts, our approach proceeds by constructing a target exiting light field consistent with the desired dual-plane constraints and then directly optimizing the lens surface shapes to realize it. This viewpoint retains the physical intuition behind prior double-freeform design methods, while the additional design flexibility offered by the two-surface geometry explicit and directly usable for joint angular--spatial regulation.

Our framework builds on the mesh-based double-freeform design pipeline of Sun et al.~\cite{sun2024mesh}, adopting its mesh representation and local ray tracing for differentiable optimization. The key conceptual advancement is that we do not treat the two target images as independent endpoints connected only through unconstrained differentiable rendering. Instead, we first establish an explicit geometric relationship between the two receptive planes via a dual-plane mapping prior computed by optimal transport. This structured prior provides per-ray guidance that regularizes the otherwise highly nonlinear, nonconvex optimization away from non-physical high-frequency local minima, and it promotes lens surfaces with larger, more physically meaningful relief variations. Moreover, because ray directions are inherently tied to how energy is transported between planes $A$ and $B$, our formulation enables explicit control over propagation directions—a capability typically absent in single-plane caustic objectives.

This capability is relevant in applications where both the pattern and direction of light are important, including advanced illumination (e.g., automotive headlamps and directional projection), industrial laser processing (where shaped beams must impinge at controlled angles), volumetric displays, structured-light 3D scanning, and optical trapping or lithography. By reducing hardware complexity while expanding controllability, our approach supports more compact and cost-effective optical systems without sacrificing performance.

\section{Method}

Our goal is to design a double-freeform lens that transforms an incident uniform collimated beam into two prescribed irradiance distributions, $g_A$ and $g_B$, on two receptive planes $A$ and $B$, respectively. 
We cast this design task as an optimization problem in which the incident and exit surface geometries are jointly adjusted so that the simulated irradiance on both planes matches the targets while satisfying the physical and geometric constraints induced by refraction and surface realizability. 
In the following, we first describe our forward light-transport model, then construct a target outgoing light field via an optimal-transport-based preprocessing step, and finally formulate the objective function and its constituent terms together with the numerical optimization procedure used to minimize the resulting energy.

\begin{figure}[h]
\centering
\includegraphics[width=0.95\linewidth]{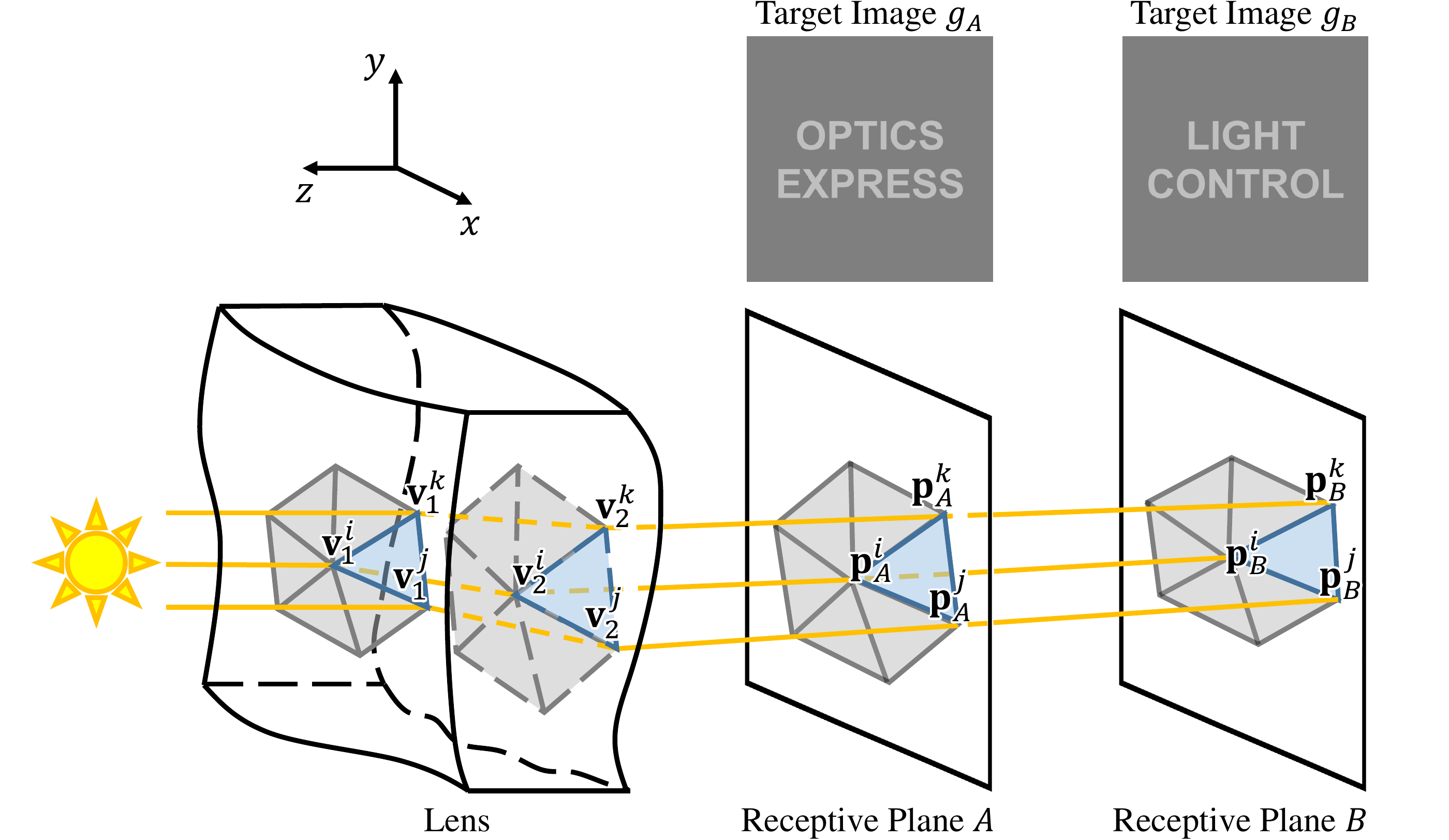}
\caption{Overview of the forward light-transport model. A collimated light beam passing through an incident-surface triangle $\triangle \mathbf{v}_1^i \mathbf{v}_1^j \mathbf{v}_1^k$ is refracted toward its corresponding exit-surface triangle $\triangle \mathbf{v}_2^i \mathbf{v}_2^j \mathbf{v}_2^k$. After the second refraction, the outgoing rays intersect receptive planes $A$ and $B$, forming imaging triangles $\triangle \mathbf{p}_A^i \mathbf{p}_A^j \mathbf{p}_A^k$ and $\triangle \mathbf{p}_B^i \mathbf{p}_B^j \mathbf{p}_B^k$, respectively. Each imaging triangle carries the same luminous flux as its corresponding incident-surface triangle. The per-pixel irradiance on each plane is obtained by accumulating the flux contributions from all overlapping imaging triangles and then converting to pixel values via gamma encoding, producing simulated images that can be directly compared with the prescribed target images $g_A$ and $g_B$.}
\label{fig:propagation}
\end{figure}

\subsection{Forward light-transport model}
We begin by discretizing both the incident and exit surfaces of the lens into triangular meshes, where three-dimensional vertices are connected by edges to form triangular facets. We apply this triangulation process identically to both surfaces, each consisting of $N$ vertices with the same connectivity. This yields two mesh surfaces whose vertices and edges are in one-to-one correspondence. We denote the $i$-th vertex on the incident surface and the corresponding vertex on the exit surface by $\mathbf{v}_1^i \in \mathbb{R}^3$ and $\mathbf{v}_2^i \in \mathbb{R}^3$, respectively, for $i \in \mathcal{S} = \{1,2,\dots,N\}$ (see Fig.~\ref{fig:propagation}). In this way, the $N$ pairs of corresponding vertices establish a bijective map between the two surfaces, which facilitates the subsequent computation of light propagation.

For an incident collimated beam with direction $\mathbf{d}_{\text{in}} = (0, 0, -1)$ and spatially uniform irradiance across the lens aperture, we connect each pair of corresponding vertices $(\mathbf{v}_1^i,\mathbf{v}_2^i)$ by a ray segment $l_{\text{mid}}^i=\overline{\mathbf{v}_1^i\mathbf{v}_2^i}$, which serves as the assumed internal propagation direction in our discrete formulation. Initially, this segment may not coincide with the physically correct refracted ray inside the lens; however, as the optimization proceeds, the integrability condition imposed on the surface normal fields (detailed in Section~\ref{sec:Optimization}) enforces agreement between the surface normals derived from $l_{\text{mid}}^i$ and those required by Snell's law at the entrance surface, so that the assumed paths progressively converge to physically valid refracted trajectories.
This formulation eliminates the need for explicit ray tracing to locate the intersection of each refracted ray with the exit surface at every iteration, thereby greatly reducing computational complexity while maintaining a consistent geometric mapping framework.

Next, we consider the refraction at the exit surface. For a given incident ray $l_{\text{mid}}^i = \overline{\mathbf{v}_1^i\mathbf{v}_2^i}$ and an auxiliary variable $\mathbf{n}_2^i$ representing the surface normal at the vertex $\mathbf{v}_2^i$, the outgoing ray $l_{\text{out}}^i$ can be computed according to Snell’s law. The intersections of $l_{\text{out}}^i$ with the receptive planes $A$ and $B$ yield the corresponding imaging points $\mathbf{p}_A^i$ and $\mathbf{p}_B^i$, respectively. 
In general, the three rays passing through the vertices of an incident-surface triangle $\triangle \mathbf{v}_1^i \mathbf{v}_1^j \mathbf{v}_1^k$ are  refracted through the corresponding vertices of the exit-surface triangle $\triangle \mathbf{v}_2^i \mathbf{v}_2^j \mathbf{v}_2^k$, and then intersect planes $A$ and $B$ at $\mathbf{p}_A^i,\mathbf{p}_A^j,\mathbf{p}_A^k$ and $\mathbf{p}_B^i,\mathbf{p}_B^j,\mathbf{p}_B^k$, respectively, forming the imaging triangles $\triangle \mathbf{p}_A^i \mathbf{p}_A^j \mathbf{p}_A^k$ and $\triangle \mathbf{p}_B^i \mathbf{p}_B^j \mathbf{p}_B^k$ (see Fig.~\ref{fig:propagation}). Assuming zero energy loss during propagation, all four triangles—on the incident surface, exit surface, and the two receptive planes—carry identical luminous flux $\Phi_{ijk}$, which is directly proportional to the projected area of the initial incident triangle $\triangle \mathbf{v}_1^i \mathbf{v}_1^j \mathbf{v}_1^k$ on the $xy$-plane.

Finally, to obtain per-pixel flux without stochastic ray sampling, we compute the exact overlap area between each imaging triangle and every pixel cell it intersects on the two receptive planes $A$ and $B$, following the area-accumulation approach for per-pixel flux computation in~\cite{sun2025end}.
For $X\in\{A,B\}$, let $P_{X,p}$ be the (axis-aligned) square region of pixel $p$ on plane $X$, and let $\triangle \mathbf{p}_X^i \mathbf{p}_X^j \mathbf{p}_X^k$ be the corresponding imaging triangle induced by the surface triangle $\triangle \mathbf{v}_1^i \mathbf{v}_1^j \mathbf{v}_1^k$. 
We assume that the flux density is uniform within $\triangle \mathbf{p}_X^i \mathbf{p}_X^j \mathbf{p}_X^k$; therefore, the contribution of this triangle to pixel $p$ is proportional to the area of their overlap. 
Specifically, we compute the intersection area
$A_{X,p}^{ijk}=\mathrm{area}\!\left(\big(\triangle \mathbf{p}_X^i \mathbf{p}_X^j \mathbf{p}_X^k\big)\cap P_{X,p}\right)$,
and accumulate the pixel flux by
$\Delta\Phi_{X,ijk\to p}=\Phi_{ijk}\,\dfrac{A_{X,p}^{ijk}}{\mathrm{area}(\triangle \mathbf{p}_X^i \mathbf{p}_X^j \mathbf{p}_X^k)}$,
yielding $\Phi_{X,p}=\sum_{(i,j,k)}\Delta\Phi_{X,ijk\to p}$. Here $\Phi_{X,p}$ denotes the flux accumulated on pixel $p$ of plane $X$. Since the pixel area is normalized to one in our implementation, this quantity is numerically equivalent to the irradiance on that pixel.
We finally map the simulated irradiance to pixel value via gamma encoding $I_{X,p}=\big(\Phi_{X,p}\big)^{1/\gamma}$ with $\gamma = 2.2$, producing the simulated images $\{I_{A,p}\}_p$ and $\{I_{B,p}\}_p$.

\subsection{Construction of the target outgoing light field}
\begin{figure}[h]
\centering
\includegraphics[width=1.0\linewidth]{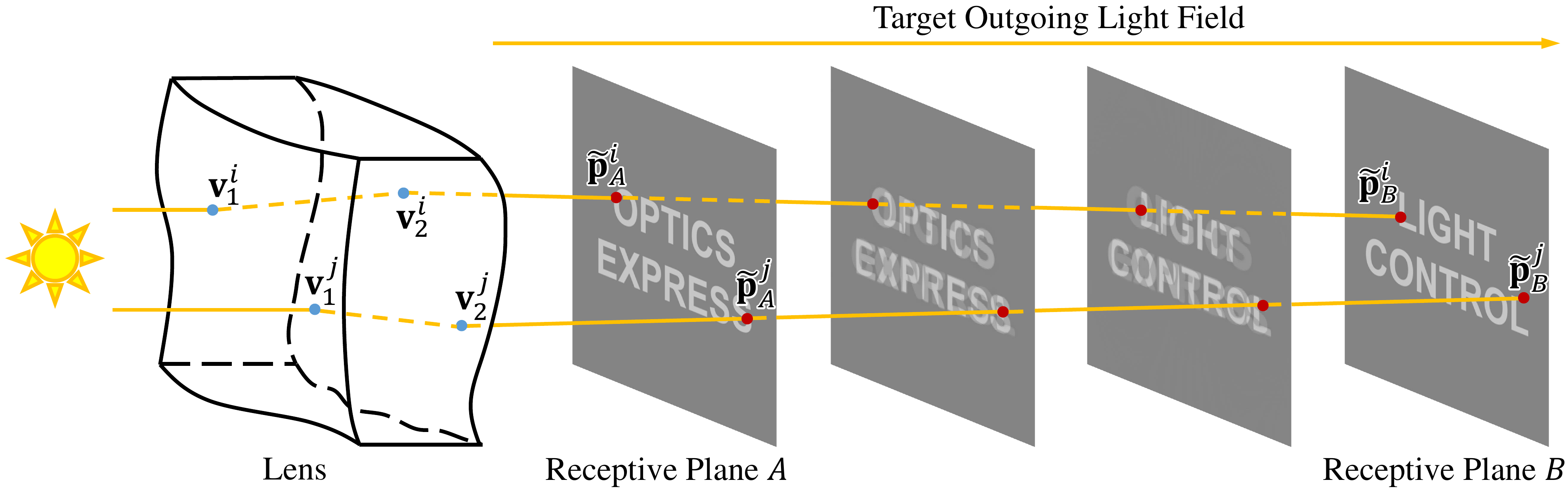}
\caption{Target outgoing light field. For each ray sample (vertex index) $i$ on the double-freeform lens, we assign a pair of OT-derived target locations $(\tilde{\mathbf{p}}_A^i,\tilde{\mathbf{p}}_B^i)$ on the two receptive planes $A$ and $B$. 
Each pair defines a target optical-path line $\tilde{l}_{\mathrm{out}}^i$ (yellow), specifying both the desired landing locations on the two planes and the corresponding propagation direction between them; these per-ray targets provide structured guidance for the subsequent surface optimization.}

\label{fig:target_field}
\end{figure}

Our goal is to realize two prescribed irradiance distributions $g_A$ and $g_B$ on two receptive planes $A$ and $B$ using a single double-freeform lens. 
Unlike single-plane caustic design, where one may construct a single correspondence between the incident distribution and a target image (e.g.,~\cite{schwartzburg2014high}), our dual-plane setting requires richer guidance: each ray must be steered so that the transported energy matches the desired allocation on both planes. 
We therefore first construct a \emph{target outgoing light field}, which in our setting denotes the angular and spatial distribution of the refracted rays after leaving the lens. The prescribed patterns on planes $A$ and $B$ are the corresponding irradiance distributions induced by this outgoing light field. Specifically, we construct it by assigning each ray sample $i$ a pair of target intersection positions $(\tilde{\mathbf{p}}_A^i,\tilde{\mathbf{p}}_B^i)$ on planes $A$ and $B$, which together define the desired propagation direction between the two planes via the target optical-path line $\tilde{l}_{\mathrm{out}}^i$. At the same time, the collection of these target intersections is constructed so that the transported energy matches the prescribed irradiance distributions on both receptive planes  (Fig.~\ref{fig:target_field}).

A proper target light field should be globally consistent with the target images while remaining geometrically well behaved. 
In particular, we prefer correspondences that induce minimal displacement from a uniform incident sampling, since highly irregular, clustered, or strongly crossing assignments would require large ray rearrangements and impose a high fitting cost on the subsequent geometric optimization. 
This consideration is especially important in our setting. Purely rendering-driven surface optimization pipelines (e.g.,~\cite{sun2024mesh}) typically start from a planar or weakly perturbed surface and minimize an image-domain discrepancy by backpropagating through refraction and rasterization. 
Such objectives provide supervision primarily at the image/distribution level and do not explicitly specify where individual rays should land; consequently, the optimizer must infer a global energy reallocation from local gradient signals. 
In a highly nonlinear and nonconvex landscape, this often makes the result sensitive to initialization and increases optimization difficulty.

\begin{figure}[h]
\centering
\includegraphics[width=1.0\linewidth]{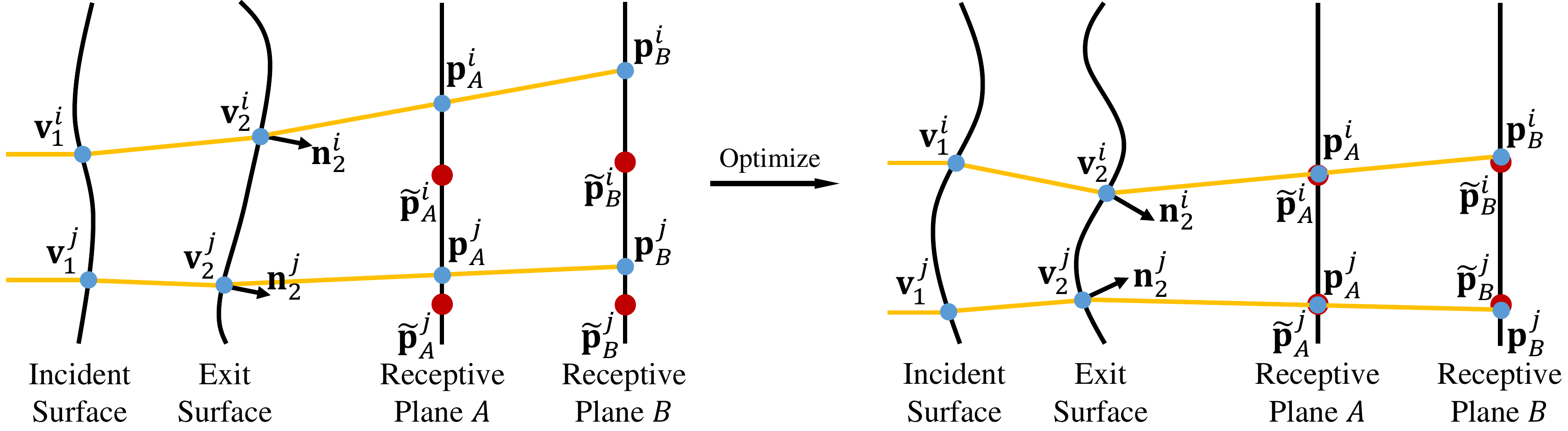}
\caption{Side view of the OT-guided optimization process. The red points \(\tilde{\mathbf{p}}_A^{i,j}\) and \(\tilde{\mathbf{p}}_B^{i,j}\) denote the fixed OT-derived target intersections on planes \(A\) and \(B\), respectively, while the blue points \(\mathbf{p}_A^{i,j}\) and \(\mathbf{p}_B^{i,j}\) denote the actual intersections of the refracted rays. During optimization, the incident and exit surfaces evolve through updates of the vertex positions \(\mathbf{v}_1^{i,j}\) and \(\mathbf{v}_2^{i,j}\), together with the exit-surface normals \(\mathbf{n}_2^{i,j}\), so that the actual ray intersections progressively approach the OT-derived targets on both receptive planes.}
\label{fig:sideview}
\end{figure}

These requirements motivate using optimal transport (OT) as a preprocessing step. In the continuous formulation, OT is closely related to Monge--Amp\`ere-type formulations; here, however, we use it purely as a discrete preprocessing tool to construct target correspondences. More specifically, OT computes a mass-preserving mapping that matches a target distribution while minimizing a global transport cost, thereby producing coherent correspondences with small, well-structured motion. 
Concretely, we solve two OT problems with a common \emph{uniform} source measure over the incident aperture. 
For each $X\in\{A,B\}$, we transport the uniform source irradiance to the target image distribution $g_X$ on plane $X$, obtaining a transport map that assigns each ray sample $i$ to a target location $\tilde{\mathbf{p}}_X^i$. 
We compute both transport maps using the Fast-OT algorithm~\cite{nader2018instant}, a fast discrete OT solver that computes an approximate transport solution rather than the exact optimum, which makes it efficient in practice. 
We perform OT independently for $A$ and $B$ (rather than transporting $g_A$ to $g_B$) for two complementary reasons. 
First, we seek correspondences that remain close to the \emph{uniform} incident sampling on each plane, yielding geometrically regular assignments with small overall motion; solving OT from the uniform source to $g_A$ and to $g_B$ directly minimizes these source-to-target transport costs, whereas transporting $g_A$ to $g_B$ does not constrain the displacement with respect to the incident parameterization.
Second, our imaging model and objective evaluate the two planes separately, so per-plane OT avoids imposing an arbitrary coupling between $g_A$ and $g_B$ during preprocessing while still producing targets consistent with each image.
Thus, we do not compute a transport map from plane $A$ to plane $B$; instead, for each ray sample $i$, the two OT solutions provide a pair of target intersections $(\tilde{\mathbf{p}}_A^i,\tilde{\mathbf{p}}_B^i)$ on planes $A$ and $B$, respectively. This pair defines the target optical-path line $\tilde{l}_{\mathrm{out}}^i$, which is used to guide the subsequent surface optimization. 
The overall pipeline is illustrated in Fig.~\ref{fig:sideview}.

\subsection{Optimization Process}
\label{sec:Optimization}

Our optimization variables are the collections $\{\mathbf{v}_1^i\}_{i\in\mathcal{S}}$ (incident-surface vertices), $\{\mathbf{v}_2^i\}_{i\in\mathcal{S}}$ (exit-surface vertices), and $\{\mathbf{n}_2^i\}_{i\in\mathcal{S}}$ (auxiliary exit-surface normals), as previously defined. 
Using the OT-derived targets $\{\tilde{\mathbf{p}}_A^i,\tilde{\mathbf{p}}_B^i\}_i$ and the associated target lines $\{\tilde{l}_{\mathrm{out}}^i\}_i$, we formulate a nonlinear optimization problem that deforms the two lens surfaces to match the desired dual-plane constraints while enforcing physical and geometric validity.
We minimize a weighted energy that combines (i) dual-plane ray-target consistency, (ii) Snell-consistent integrable normal fields, (iii) incident-side mass (projected-area) consistency with the OT source measure, and (iv) a geometric barrier to prevent invalid surface configurations. 
The individual terms are introduced below and finally combined into a single objective.

\paragraph{Dual-plane ray-target consistency.}
We require that the outgoing ray originating from the exit-surface vertex $\mathbf{v}_2^i$ sequentially pass through the target points $\tilde{\mathbf{p}}_A^i$ and $\tilde{\mathbf{p}}_B^i$ on the two receptive planes, where $\tilde{\mathbf{p}}_A^i$ and $\tilde{\mathbf{p}}_B^i$ are obtained from an optimal-transport (OT) preprocessing step described after the formulation. 
Accordingly, we impose two complementary penalties: (1) $\mathbf{v}_2^i$ should lie close to the target optical-path line $\tilde{l}_{\text{out}}$ defined by $\tilde{\mathbf{p}}_A^i$ and $\tilde{\mathbf{p}}_B^i$, and (2) the actual intersection $\mathbf{p}_B^i$ on plane $B$ should match $\tilde{\mathbf{p}}_B^i$. 
This yields a distance term
\begin{equation}
    E_{\text{dist}} = \sum\nolimits_{i \in \mathcal{S}} \mathrm{dist}(\mathbf{v}_2^i, \tilde{l}_{\text{out}})^2,
\end{equation}
and an alignment term
\begin{equation}
    E_{\text{align}} = \sum\nolimits_{i \in \mathcal{S}} \|\mathbf{p}_B^i - \tilde{\mathbf{p}}_B^i\|^2.
\end{equation}
Here $\mathrm{dist}(\cdot)$ denotes the point-to-line distance and $\mathcal{S}$ is the set of mesh vertices, where each vertex index $i\in\mathcal{S}$ corresponds to one ray sample used in the OT preprocessing.

\paragraph{Snell-consistent integrability.}
To ensure physical realizability of the two surfaces, we enforce Snell’s law compliance at both the incident and exit surfaces by constraining their derived unit normal fields $\{\overline{\mathbf{n}}_1^i\}$ and $\{\overline{\mathbf{n}}_2^i\}$ to satisfy an integrability condition. 
We first prescribe the incoming direction as
$\mathbf{d}_{\text{in}}=(0,0,-1)$
and the intermediate refracted direction as
$\mathbf{d}_{\text{mid}}^i
=
\frac{\mathbf{v}_2^i-\mathbf{v}_1^i}{\|\mathbf{v}_2^i-\mathbf{v}_1^i\|}$.
For this pair of directions, the incident-side unit normal \(\overline{\mathbf{n}}_1^i\) is obtained analytically from the vector form of Snell's law:
\[
\overline{\mathbf{n}}_1^i
=
\frac{\eta \mathbf{d}_{\text{mid}}^i-\mathbf{d}_{\text{in}}}
{\|\eta \mathbf{d}_{\text{mid}}^i-\mathbf{d}_{\text{in}}\|},
\]
where \(\eta \approx 1.49\) is the relative refractive index. This formula is equivalent to Eq.~(2.1) in~\cite{gutierrez2013refractor}.
For the exit surface, we use the normalized auxiliary variable $\overline{\mathbf{n}}_2^i=\frac{\mathbf{n}_2^i}{\|\mathbf{n}_2^i\|}$. 
Following~\cite{sun2016vertex}, we adopt the integrability term
\begin{equation}
E_{\mathrm{int}} = \sum\nolimits_{(i, j) \in \mathcal{E}} \frac{\overline{\mathbf{e}}_1 \cdot (\overline{\mathbf{n}}_1^i + \overline{\mathbf{n}}_1^j)}{\|\overline{\mathbf{n}}_1^i + \overline{\mathbf{n}}_1^j\|} + \lambda \frac{\overline{\mathbf{e}}_2 \cdot (\overline{\mathbf{n}}_2^i + \overline{\mathbf{n}}_2^j)}{\|\overline{\mathbf{n}}_2^i + \overline{\mathbf{n}}_2^j\|},
\label{Int_Loss}
\end{equation}
where $\mathcal{E}$ is the edge set, $\overline{\mathbf{e}}_1$ and $\overline{\mathbf{e}}_2$ are unit edge vectors, and $\lambda$ is a user-defined parameter. 
Intuitively, this term encourages the average normal across each edge to be orthogonal to the edge direction, promoting an integrable normal field.

\paragraph{Incident-side mass consistency.}
Our OT preprocessing is solved on a fixed discretization of the incident aperture under uniform illumination: each incident-surface triangle $\triangle \mathbf{v}_1^i \mathbf{v}_1^j \mathbf{v}_1^k$ carries a flux weight proportional to its projected area on the $xy$-plane, i.e., $\Phi_{ijk}\propto A(\triangle \mathbf{v}_1^i \mathbf{v}_1^j \mathbf{v}_1^k)$. 
The OT targets are therefore computed with respect to this \emph{fixed} discretized source measure. 
During surface fitting, if the projected area of an incident triangle changes significantly, its associated flux weight changes accordingly, which alters the source measure and makes the OT-derived correspondences inconsistent. 
To maintain consistency with the OT construction throughout optimization, we penalize deviations of each triangle's projected area from its initialization $A^0_{ijk}$:
\begin{equation}
\label{eq:mass_consistency}
E_{\text{area}}=\sum\nolimits_{\triangle \mathbf{v}_1^i \mathbf{v}_1^j \mathbf{v}_1^k\in\mathcal{F}_1}\left(A(\triangle \mathbf{v}_1^i \mathbf{v}_1^j \mathbf{v}_1^k)-A^0_{ijk}\right)^2.
\end{equation}
Here $\mathcal{F}_1$ contains all incident-surface triangles, $A(\cdot)$ denotes the signed projected area on the $xy$-plane, and $A^0_{ijk}$ is the initial projected area of triangle $\triangle \mathbf{v}_1^i \mathbf{v}_1^j \mathbf{v}_1^k$ used to define the OT source measure.

\paragraph{Geometric barrier.} 
To prevent degenerate or flipped triangles during optimization, we define

\begin{equation}
E_{\text{barr}}
=
\sum\nolimits_{\triangle \mathbf{v}_1^i \mathbf{v}_1^j \mathbf{v}_1^k\in\mathcal{F}_1}
f_{\text{barr}}\!\left(A(\triangle \mathbf{v}_1^i \mathbf{v}_1^j \mathbf{v}_1^k)\right)
+
\sum\nolimits_{\triangle \mathbf{v}_2^i \mathbf{v}_2^j \mathbf{v}_2^k\in\mathcal{F}_2}
f_{\text{barr}}\!\left(A(\triangle \mathbf{v}_2^i \mathbf{v}_2^j \mathbf{v}_2^k)\right),
\end{equation}
where
\begin{equation}
f_{\text{barr}}(a)=
\begin{cases}
\displaystyle \frac{1}{a}, & a>0,\\[0.8ex]
+\infty, & a\leq 0.
\end{cases}
\end{equation}
Here $\mathcal{F}_2$ denotes the set of all exit-surface triangles. This barrier penalizes excessively small triangles and prevents triangle flips by restricting the step size during line search.

\paragraph{Total objective.}
The total energy is a weighted sum of the above terms:
\begin{equation}
    E = w_1  E_{\text{dist}} + w_2 E_{\text{align}} + w_3 E_{\text{int}} + w_4 E_{\text{area}} + w_5 E_{\text{barr}}.
\label{Total_Loss}
\end{equation}
Here $w_1,\ldots,w_5$ are user-specified weighting parameters that balance the contributions of the above terms.

\subsection{Implementation Details}
We minimize Eq.~\eqref{Total_Loss} using the limited-memory BFGS (L-BFGS) algorithm~\cite{liu1989limited}. 
Unless otherwise specified, we use fixed weights $(w_1,w_2,w_3,w_4,w_5)=(10^5,10^9,10^3, 10^4, 10^{-5})$ and $\lambda=10^{-2}$ in Eq.~\eqref{Int_Loss} for all experiments. These values were chosen empirically through preliminary experiments to balance the numerical scales and relative roles of the loss terms. In particular, $w_2$ is set relatively large because the corresponding term directly affects the ray--plane intersection positions and therefore influences the irradiance distributions. $w_5$ is kept small because the geometric barrier is only intended to prevent invalid configurations. The choice $\lambda=10^{-2}$ places more emphasis on the incident-surface term in the Snell-consistent integrability loss, since errors on the incident surface change where rays reach the exit surface; if rays do not arrive at the intended exit-surface vertices, enforcing Snell's law there becomes less meaningful.

The target images are specified at a resolution of $256\times256$ pixels, which provides sufficient spatial sampling for accurate light field reconstruction while maintaining computational efficiency in our optimization framework. We initialize the lens as a rectangular cuboid with a 10 cm base length and a thickness of 2 cm, where the boundary vertex coordinates along the $x-$ or $y-$ axis are fixed to maintain both the lens shape and the total incident luminous flux. 

The location range of the two receptive planes $A$ and $B$ is pattern-dependent, as significant differences between the desired irradiance distributions $g_A$ and $g_B$ require sufficient propagation distance for the light field to transform. Furthermore, plane $A$ must be positioned sufficiently close to the exit surface.
Excessive separation would lead to physically unrealistic surface geometries when reconstructing the exit surface profile through backward ray tracing (by intersecting the extensions of $\tilde{\mathbf{p}}_B^i\tilde{\mathbf{p}}_A^i$ segments with the exit surface), inevitably resulting in triangle flipping that makes the optimization problem infeasible.

For the basic case involving the text transformation from "OPTICS EXPRESS" to "LIGHT CONTROL", we position the receptive planes $A$ and $B$ at 2 cm and 10 cm from the initial exit surface, respectively. For the more complex pattern transformation from the word "Butterfly" to an actual butterfly image, the planes are placed at 1 cm and 17 cm respectively to accommodate the increased optical modulation complexity and required light field evolution distance.

\begin{figure}[t!]
\centering
\includegraphics[width=1.0\linewidth]{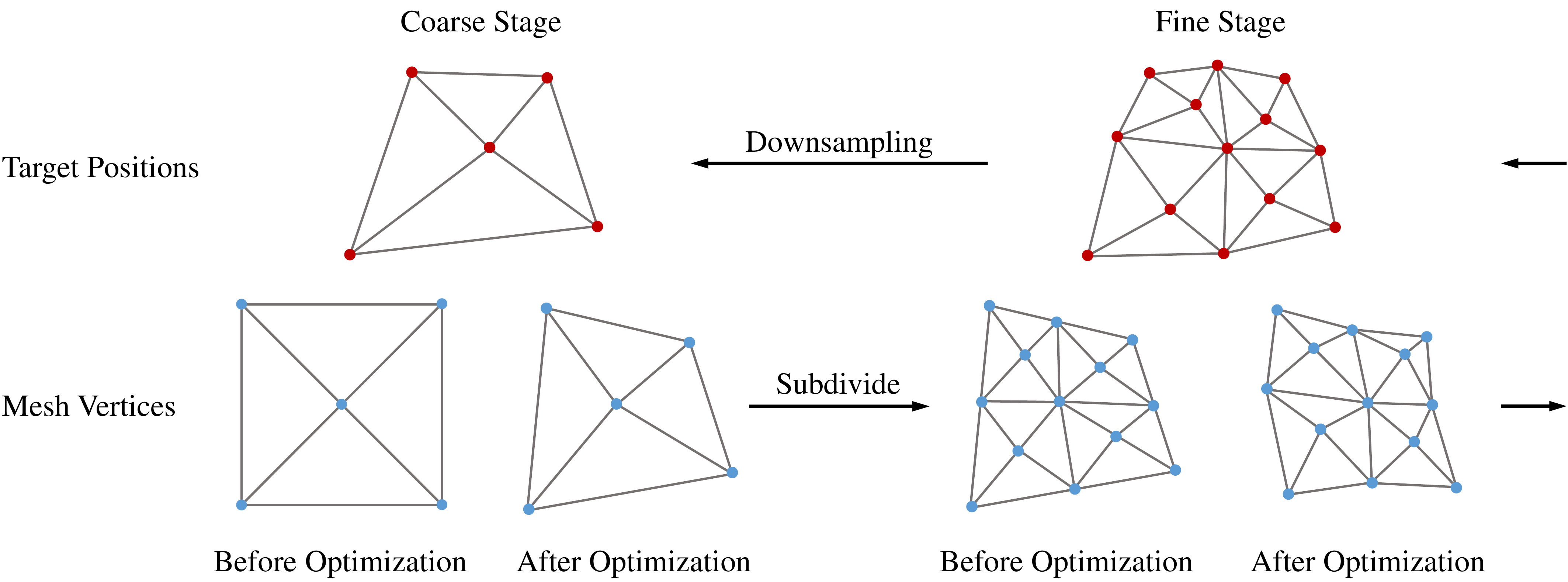}
\caption{A local example of the first two stages. The target positions are obtained by downsampling from the finest-stage, while the mesh for the finer stage is initialized by subdividing the optimized mesh from the coarser stage.}
\label{fig:Coarse_To_Fine}
\end{figure}

\begin{figure}[p!]
\centering
\includegraphics[width=0.95\linewidth]{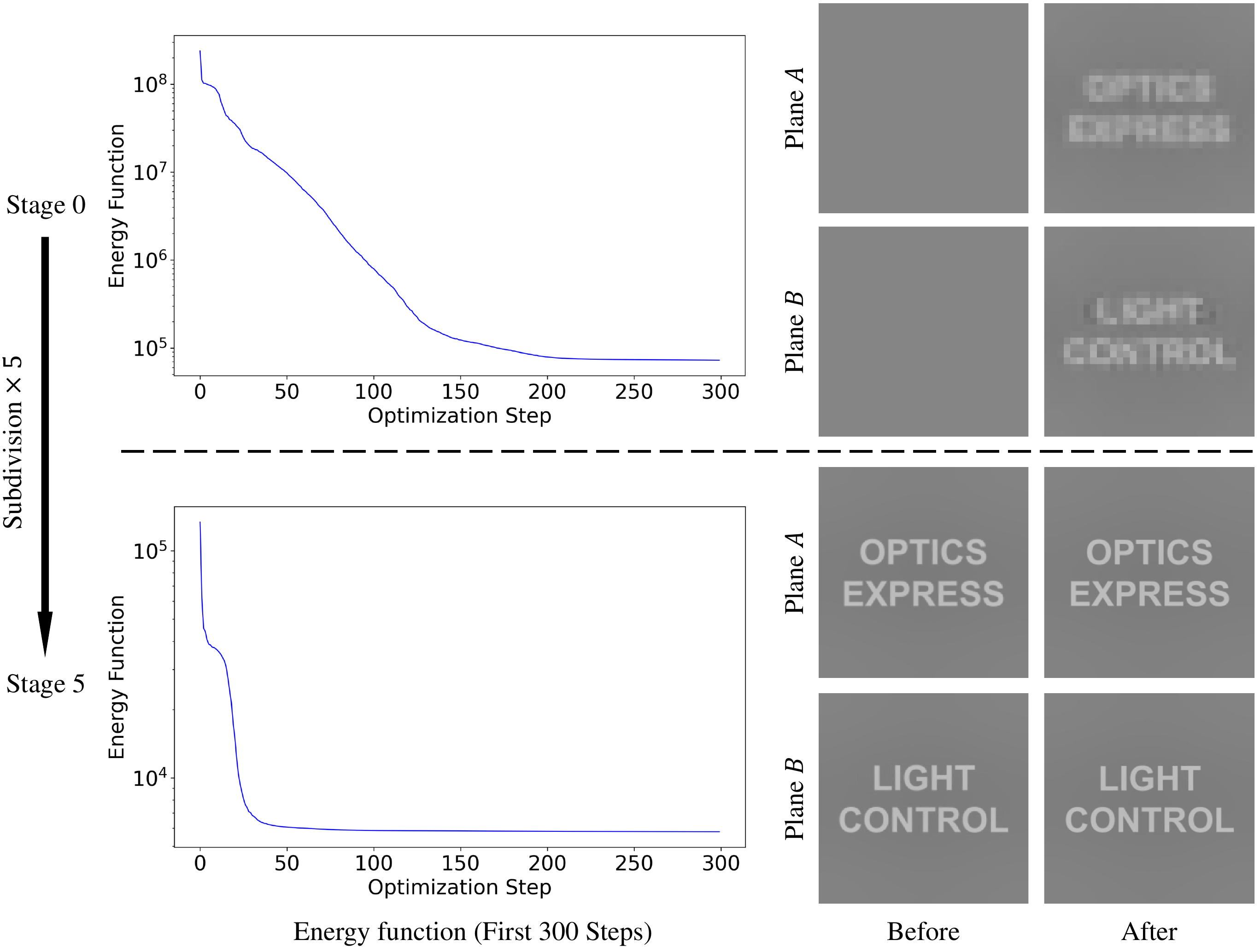}
\caption{Optimization progress across stages: the left panels show the energy reduction over the first 300 iterations at the coarsest and finest stages, while the right panels compare the simulated images on planes A and B before and after optimization for each stage.}
\label{fig:Loss}
\end{figure}

We adopt a coarse-to-fine optimization strategy to improve optimization stability and efficiency. The main idea is to first optimize a coarse mesh to capture the overall lens shape, and then progressively refine the mesh resolution and optimize its geometry to recover finer surface details. Although optimization starts from the coarsest stage, we first construct the finest-stage discretization so that all stages are derived consistently from a common reference. Specifically, we solve Fast-OT on a regular $1024\times1024$ quadrilateral mesh representing the incident aperture, which yields two sets of target positions $\{\tilde{\mathbf{p}}_A^i\}_i$ and $\{\tilde{\mathbf{p}}_B^i\}_i$ on planes $A$ and $B$, respectively. We then insert a weighted barycenter into each quad, thereby converting both the finest-stage mesh and the corresponding target positions into a triangular discretization. Starting from this finest-stage data, we repeatedly downsample the mesh vertices and the target positions to construct progressively coarser stages until reaching the coarsest stage. In this way, each stage uses a subset of the vertices and target positions defined at the finest resolution. After convergence at a stage, the optimized mesh is subdivided to initialize the next finer stage: for each quadrilateral patch triangulated into four triangles, we insert the midpoints of every edge and connect them according to the pattern shown in Fig.~\ref{fig:Coarse_To_Fine}, thereby splitting one such patch into four quadrilateral patches. This finer stage is then optimized against the corresponding denser subset of target positions. Fig.~\ref{fig:Coarse_To_Fine} shows a local example of the downsampling/subdivision process between the first two stages. In this way, the global lens shape is determined first at low resolution and then progressively refined at higher resolutions. Fig.~\ref{fig:Loss} demonstrates the progressive improvement of the simulated images across stages: the upper-right subplot shows the result after coarse-stage optimization on both receptive planes, and the lower-right subplot shows the subsequent refinement at the finest stage.

\begin{figure}[p!]
\centering
\includegraphics[width=0.95\linewidth]{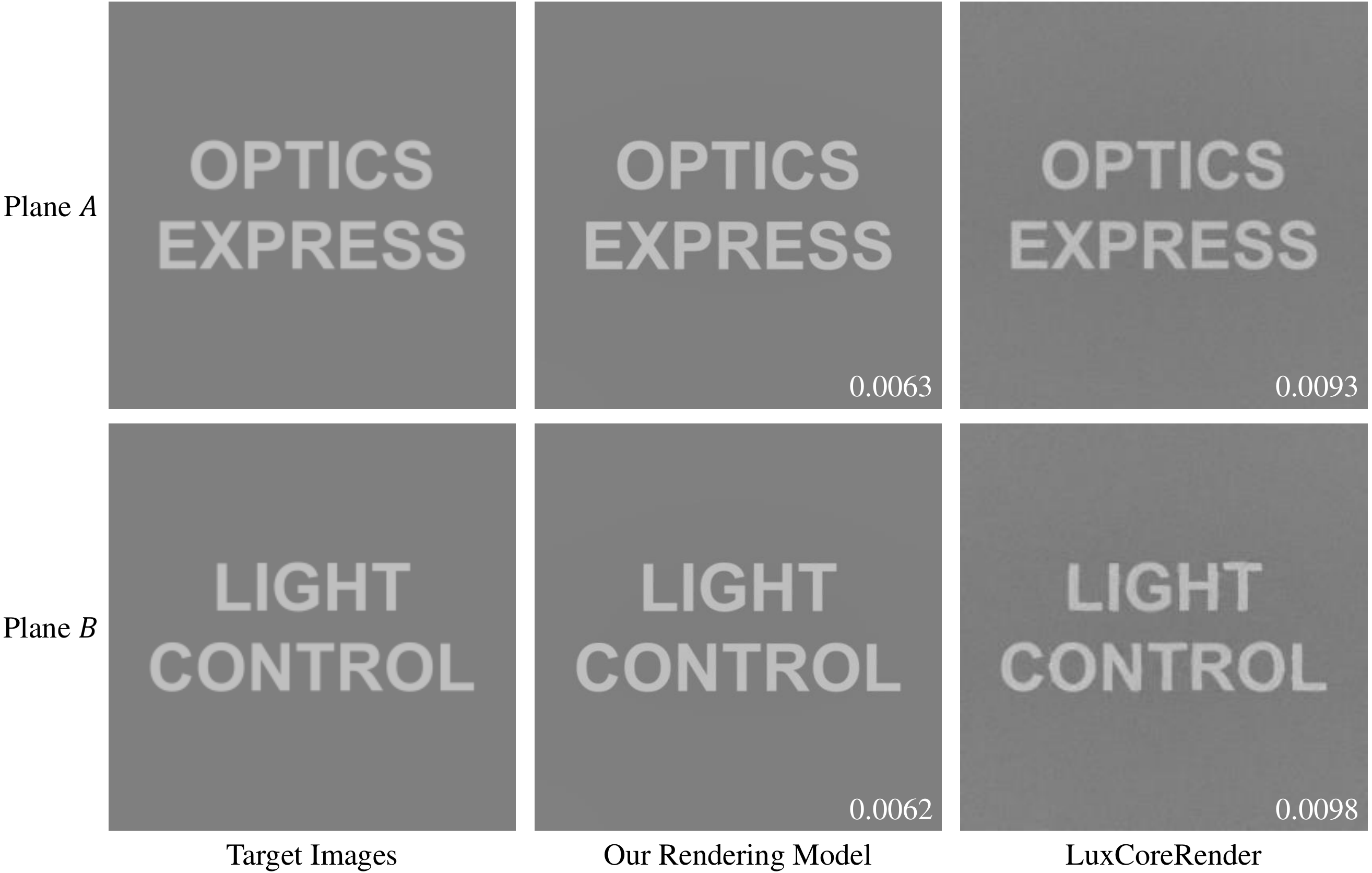}
\caption{Accuracy of our propagation model. The numerical value in the lower right corner of each image indicates the MAE with the target image.}
\label{fig:Blender}
\end{figure}

All optimizations are performed on a single NVIDIA GeForce RTX 4090 with 24GB VRAM. In the coarse-to-fine scheme, the maximum number of iterations at each stage is set to 20{,}000. Overall, optimizing the incident and exit surfaces of the lens takes about one hour, with the highest-resolution stage accounting for approximately 50 minutes.

\section{Experiments}

In this section, we comprehensively evaluate our algorithm through multiple perspectives. First, Fig.~\ref{fig:Blender} demonstrates the accuracy of our propagation model described in the previous section. We compare our simulation results with those generated by \luxcorerender\cite{luxcorerender_website}, a physically-accurate ray-tracing engine. While the \luxcorerender\ output for Plane $B$ exhibits faint blur, attributable to minor discrepancies in surface normal estimation between our auxiliary variables and \luxcorerender's refraction calculations, the overall irradiance distributions demonstrate remarkable consistency. 
Notably, the rays that form the letter `O' on Plane $A$ become fully merged into the background after propagating to Plane $B$, leaving no visible residual contour of the letter `O'. This behavior provides further evidence for the physical accuracy of our light transport model and validates that the designed lens correctly redistributes energy between distinct spatial patterns across the two planes.

\begin{figure}[!t]
\centering
\includegraphics[width=0.98\linewidth]{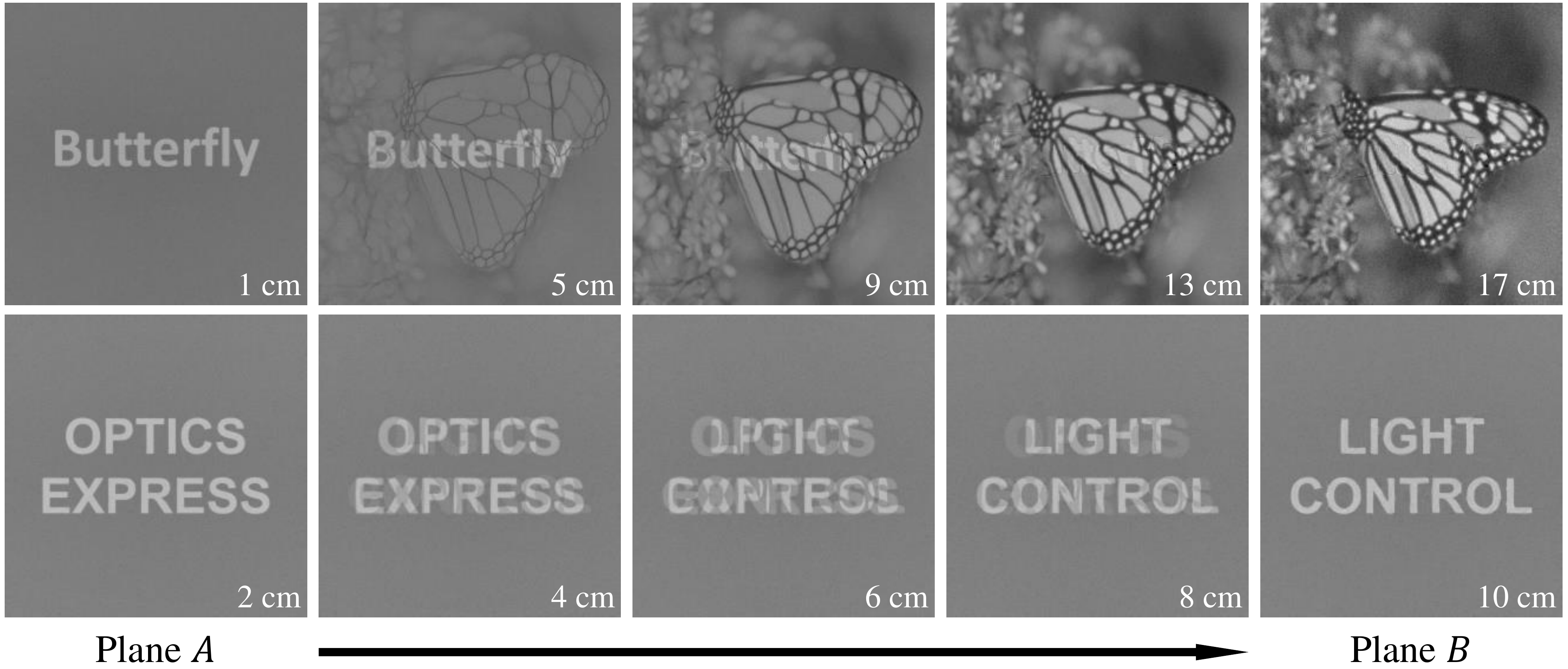}
\caption{Light field evolution from Plane $A$ to Plane $B$. The intermediate images illustrate the progressive transformation of the light field at three equidistant planes located at 25\%, 50\%, and 75\% of the propagation distance between the two planes.}
\label{fig:A_to_B}
\end{figure}

\begin{figure}[!b]
\centering
\includegraphics[width=0.98\linewidth]{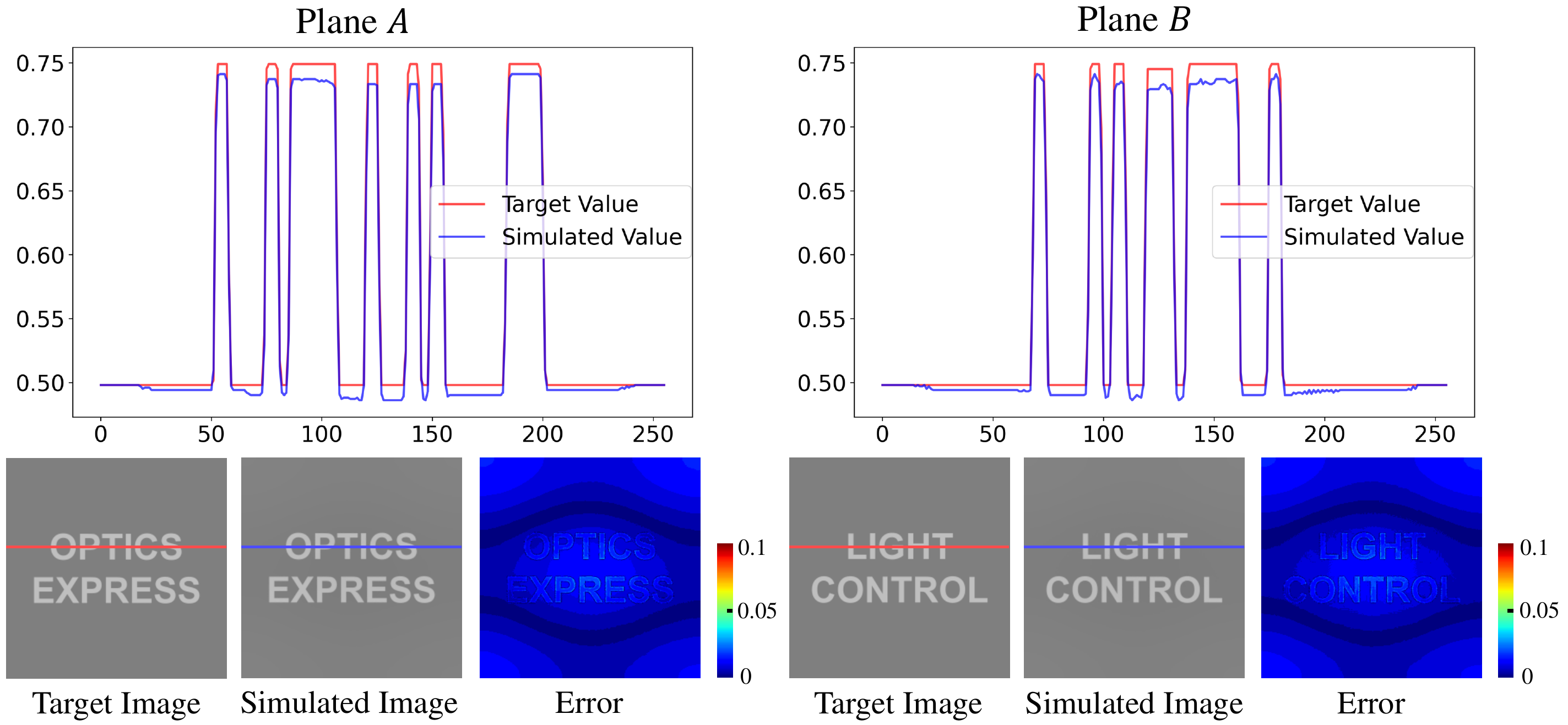}
\caption{Target and simulated irradiance distributions on planes \(A\) and \(B\). The top row compares one-dimensional profiles sampled along the indicated horizontal lines. The bottom row shows the corresponding target images, simulated images, and pixel-wise error maps.}
\label{fig:Colorbar}
\end{figure}

In the lower-right corner of each image in Fig.~\ref{fig:Blender}, we also show a quantitative measure between the simulated and target images using the \emph{Mean Absolute Error} (MAE), which is defined as:
\begin{equation}
\mae = \frac{1}{n_p}\sum_{i=1}^{n_p} \frac{|p^i - \tilde{p}^i|}{I_{\text{max}}}
\end{equation}
where $n_p$ denotes the total pixel count, $p^i$ and $\tilde{p}^i$ represent the $i$-th pixel values in the simulated and target images respectively, and $I_{\max}=255$ is the maximum possible pixel value that normalizes the error to the range $[0,1]$. The MAE values show close agreement between both our propagation model and reference ray tracing simulations with the target images, quantitatively validating the accuracy of our approach.

Fig.~\ref{fig:A_to_B} further demonstrates the continuous evolution of the light field as it propagates between receptive Planes $A$ and $B$, capturing the spatial redistribution of irradiance while maintaining energy conservation throughout the optical path. The visualization confirms our method’s ability to precisely control the light field transformation between the two planes.

\begin{figure}[!t]
\centering
\includegraphics[width=0.97\linewidth]{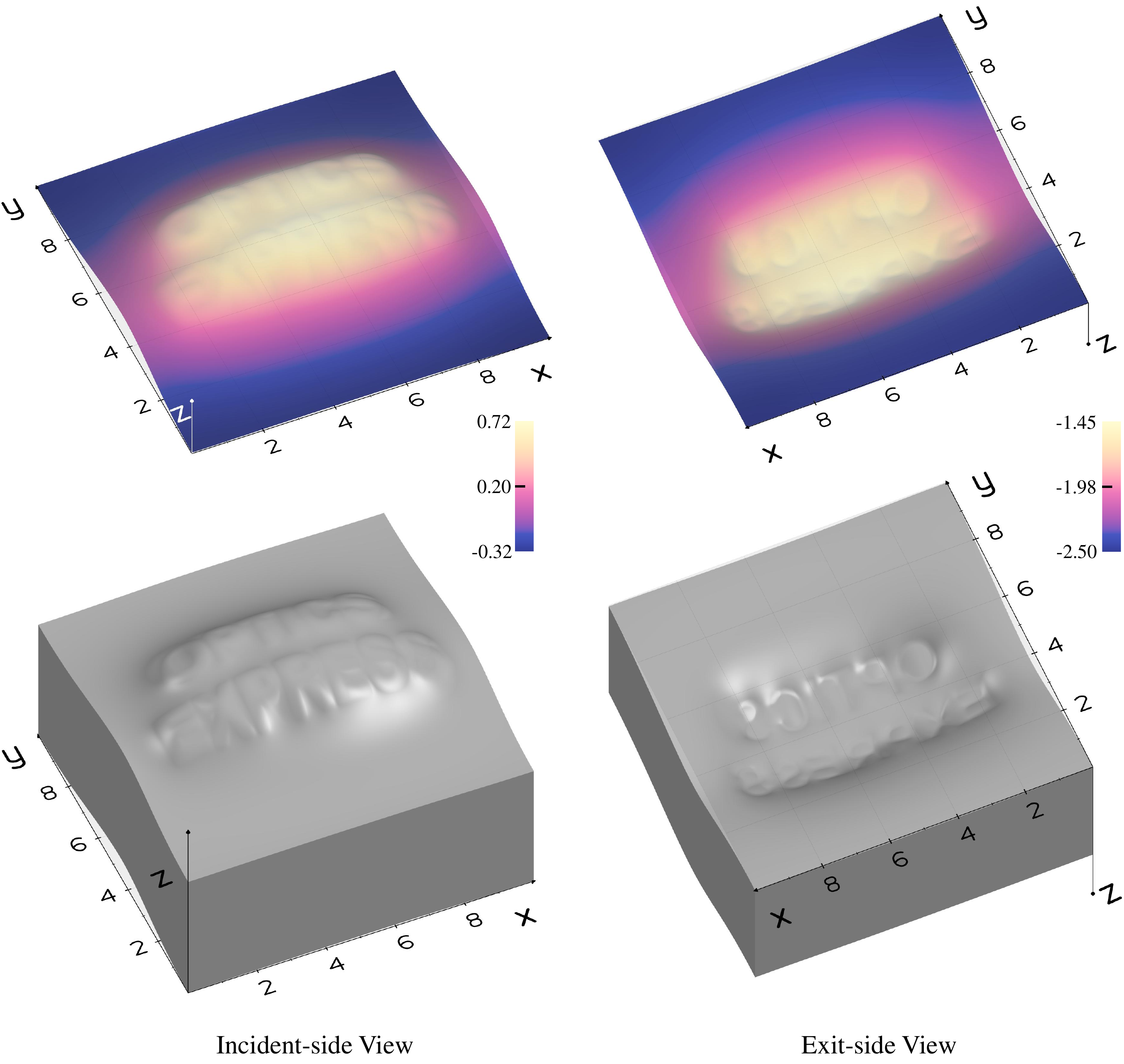}
\caption{Optimized geometries of the incident and exit freeform surfaces of the designed double-freeform lens. The top row shows the surface height profiles, where the colormaps indicate the relative height along the optical axis $z$. The bottom row shows the corresponding full lens geometries from the same viewing directions, providing a more complete visualization of the optimized incident and exit surfaces.}
\label{fig:Surface}
\end{figure}

Fig.~\ref{fig:Colorbar} shows the simulated irradiance distributions on two target planes. For both planes, the simulated irradiance distributions exhibit excellent agreement with the target patterns: “OPTICS EXPRESS” for Plane $A$ and “LIGHT CONTROL” for Plane $B$, accurately reproducing the overall irradiance shapes and fine-scale contrast details. The cross-sectional profiles show that the simulated curves closely follow the target values, with slight deviations observed in certain regions. These discrepancies arise because the irradiance distributions on Plane $A$ and Plane $B$ are difficult to satisfy simultaneously. The optimization therefore yields a physically realizable solution that minimizes the global fitting error in a "least-squares" sense. The error maps confirm that the residual differences remain small (mostly below 0.05), indicating that the proposed double-freeform design effectively balances the dual-plane irradiance objectives. Overall, the results verify that the optimized lens can reproduce distinct target irradiance distributions on separate axial planes, demonstrating precise and stable angular-spatial light-field control through a single optical element.

Fig.~\ref{fig:Surface} further demonstrates the shape of the lens. Both the incident and exit surfaces exhibit smoothly varying freeform features that encode the complex angular-spatial mapping required to project the prescribed patterns on the two target planes. Since Plane $A$ is positioned closer to the lens and is designed to form the “OPTICS EXPRESS” pattern first, subtle structural traces of that pattern can be discerned on both the incident and exit surfaces. These features locally modulate the optical path length to guide the rays into the correct spatial configuration. More importantly, through the precise optimization of the shape and thickness of the lens, we can also manipulate the propagation direction of the rays such that, after further propagation, the light reconstructs the second pattern “LIGHT CONTROL” on Plane $B$. This visualization provides an intuitive understanding of how the double-freeform geometry encodes both directional and spatial control, revealing the physical mechanism by which the dual-plane constraints are simultaneously satisfied.

\begin{figure}[!t]
\centering
\includegraphics[width=1.0\linewidth]{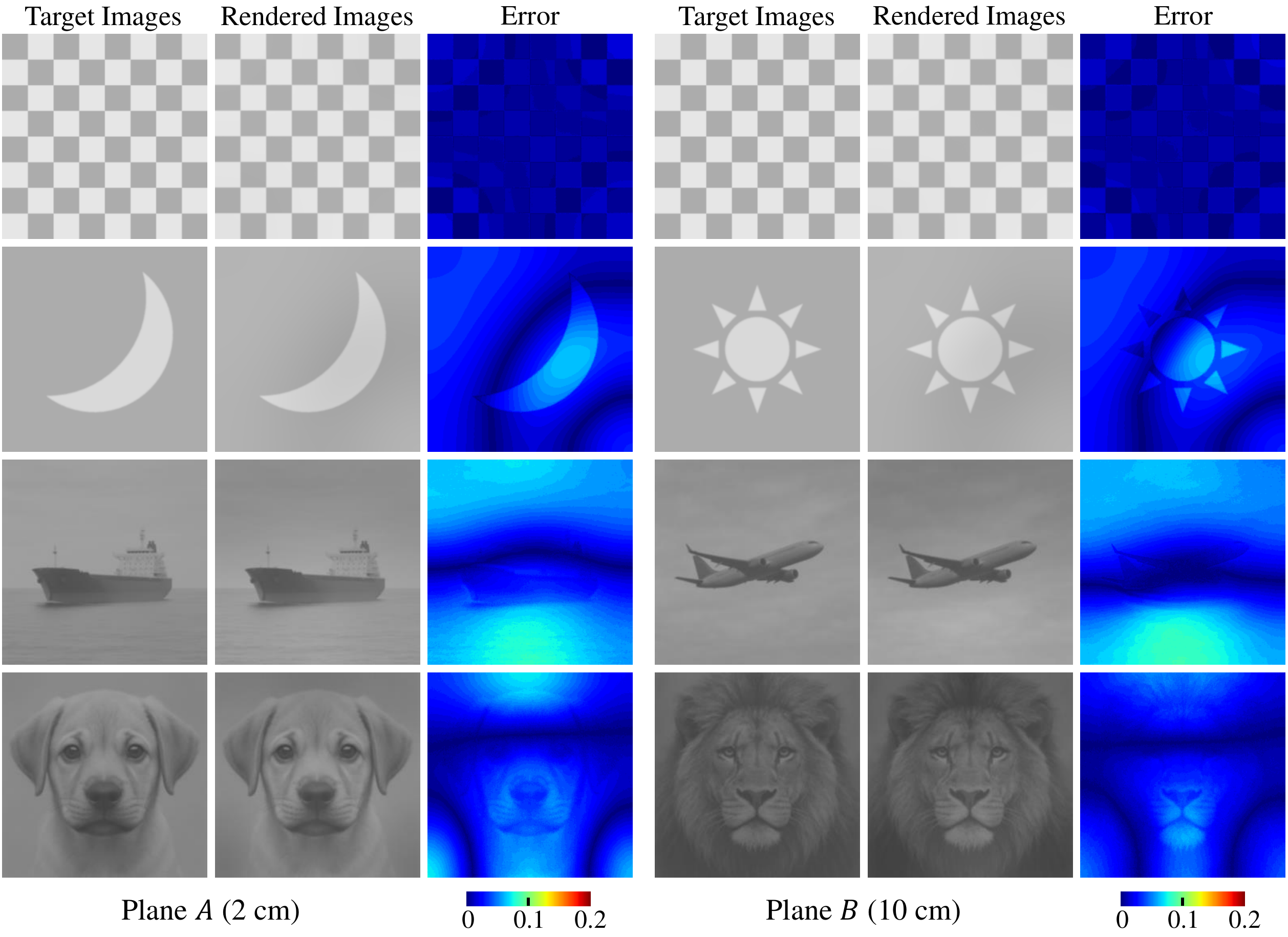}
\caption{Additional simulation examples demonstrating the effectiveness of the proposed method. The optimized double-freeform lens accurately reproduces two distinct target images on Plane $A$ and Plane $B$, respectively.}
\label{fig:Examples}
\end{figure}

\begin{figure}[!t]
\centering
\includegraphics[width=0.96\linewidth]{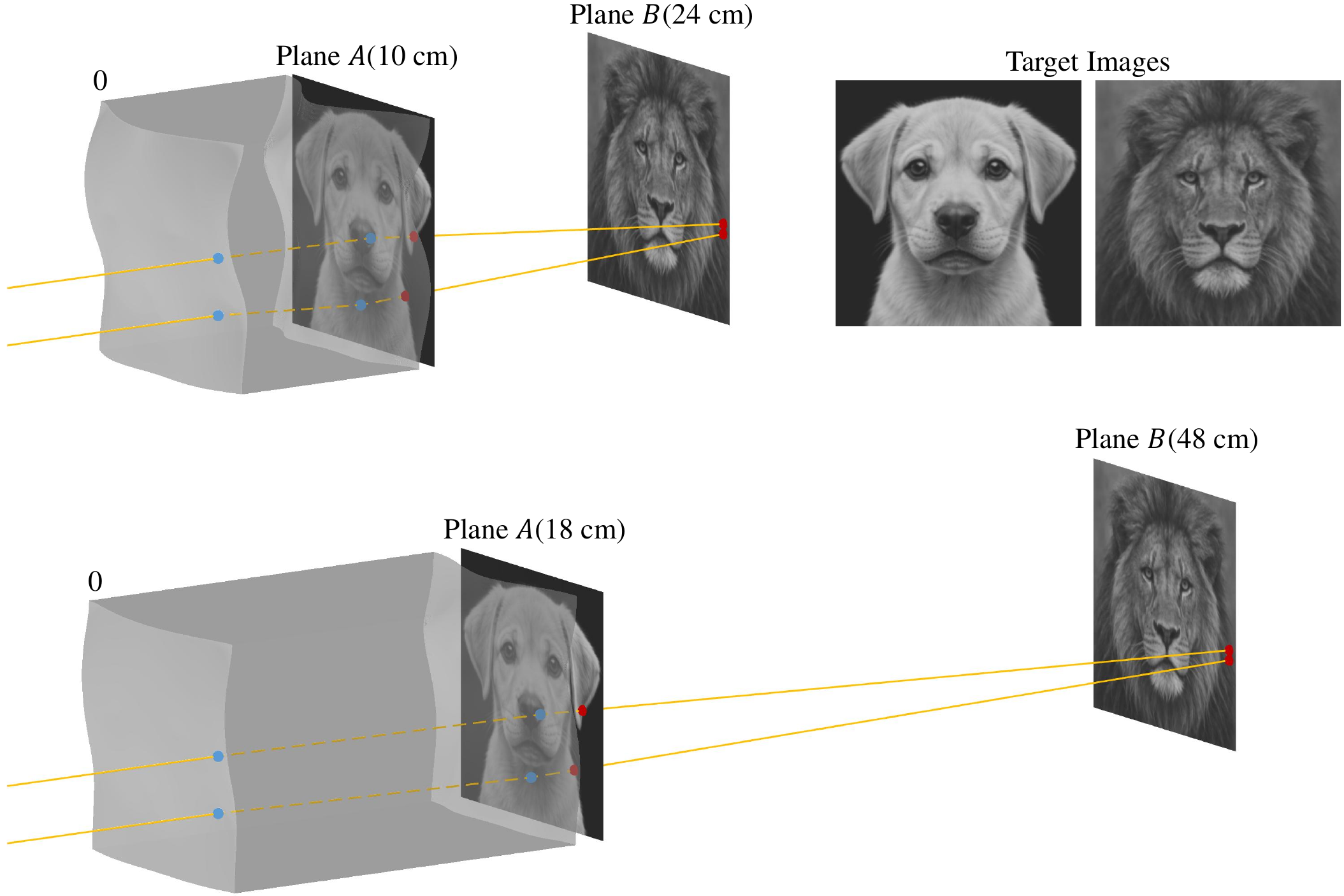}
\caption{High-contrast dual-plane design example. The target irradiance distributions are shown in the upper right. In each row, the perspective view shows the designed double-freeform lens together with representative ray paths, while the images on planes $A$ and $B$ are the simulated irradiance results. The distances indicate the positions of planes $A$ and $B$ relative to the reference position $0$, for a lens with side length $10\,\mathrm{cm}$.}
\label{fig:High_Contrast}
\end{figure}

\begin{figure}[!htbp]
\centering
\includegraphics[width=0.96\linewidth]{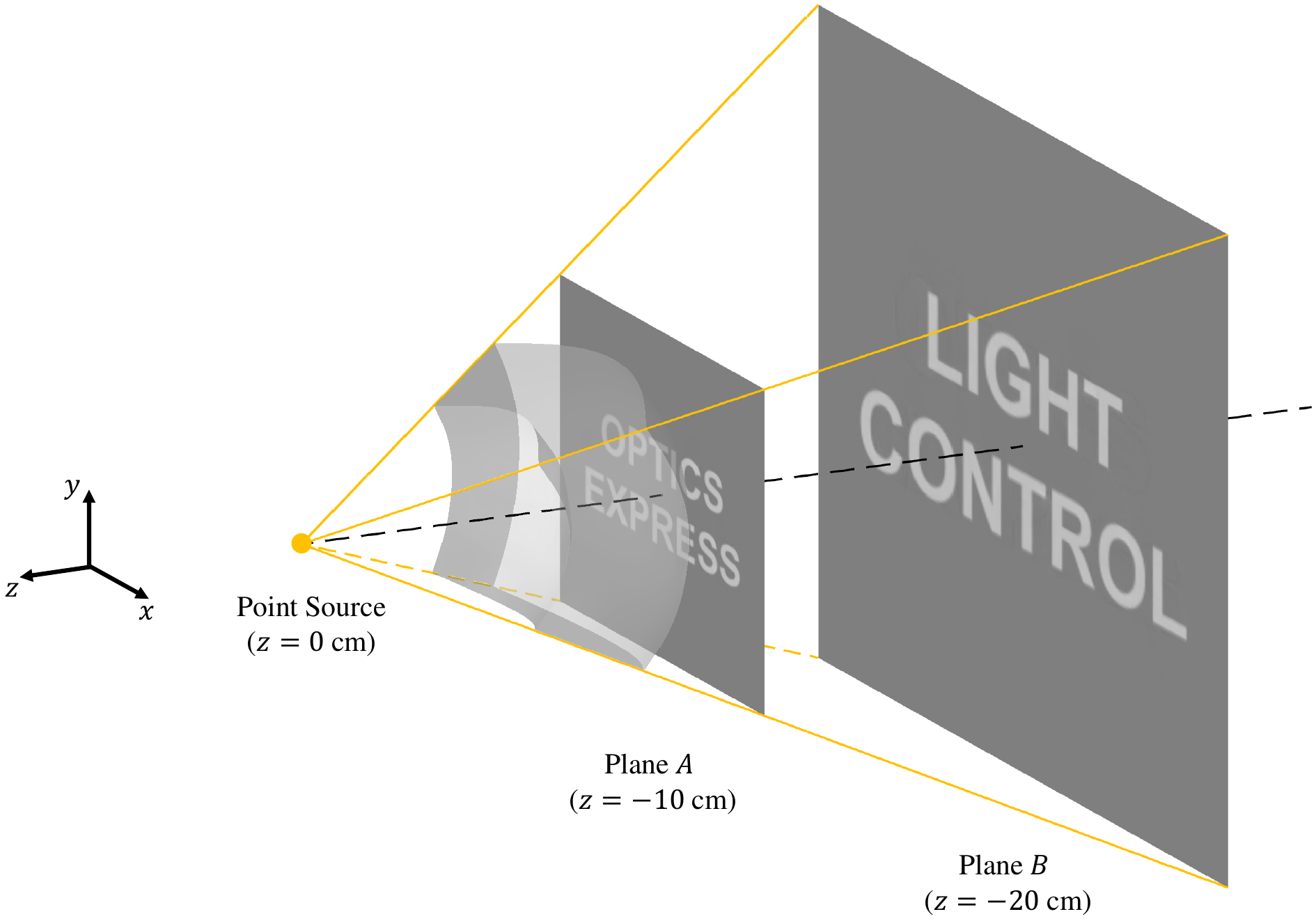}
\caption{Point-source extension of our method. The black dashed line indicates the optical axis. The yellow rays form a quadrangular pyramid that defines the propagation range, and the angle between the yellow boundary rays and the optical axis is set to $40^\circ$. The receptive planes are placed at $z=-10\,\mathrm{cm}$ and $z=-20\,\mathrm{cm}$, respectively, and the images shown on the planes are the simulated irradiance results.}
\label{fig:Point_Source}
\end{figure}

\begin{figure}[!htbp]
\centering
\includegraphics[width=1.0\linewidth]{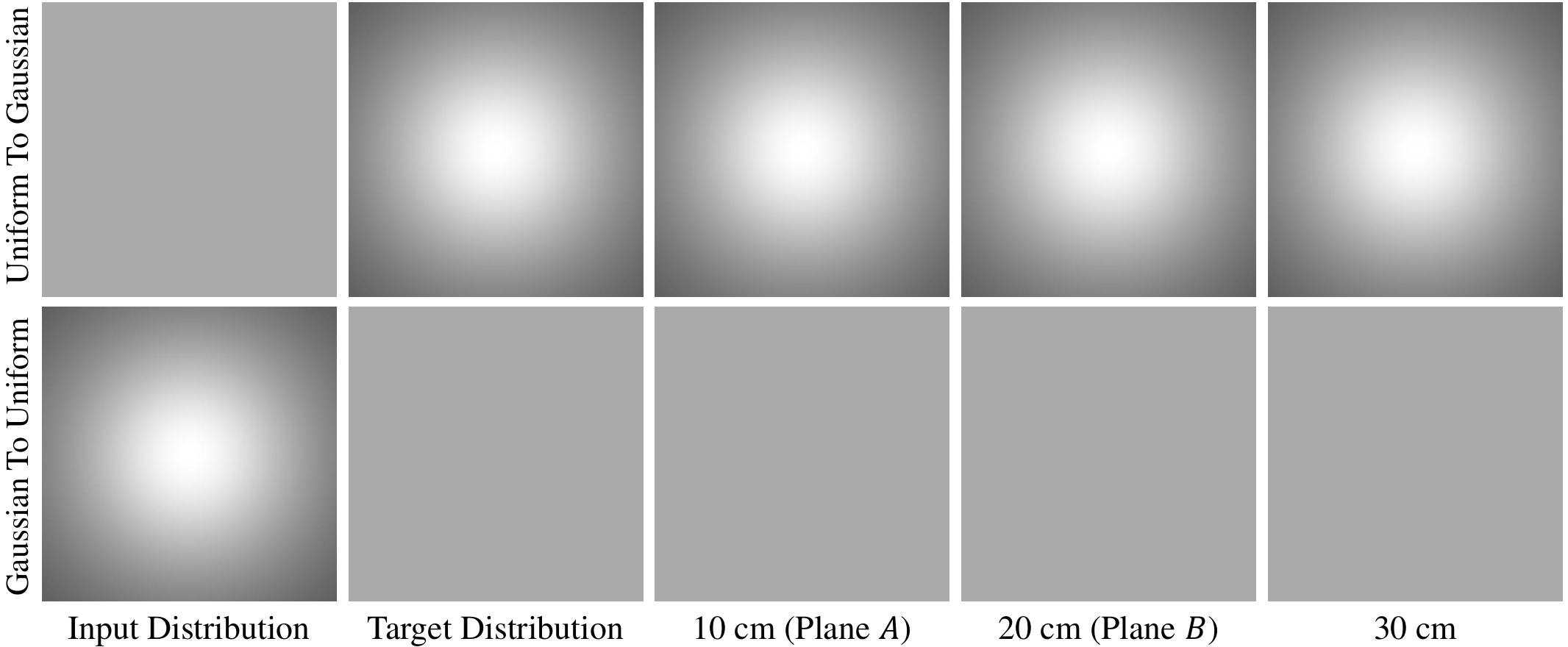}
\caption{Demonstration of bidirectional beam shaping using the proposed double-freeform lens. \textbf{Top row:} the lens is optimized to transform a uniform collimated beam into a Gaussian irradiance distribution while maintaining collimation during propagation. \textbf{Bottom row:} without re-optimization, the same lens is directly used in reverse to convert an incident Gaussian beam back into a uniform collimated beam. Together, these results demonstrate the optical reversibility and physical consistency of the proposed design.}
\label{fig:Gaussian}
\end{figure}

We further conducted additional experiments using multiple target image pairs to validate the robustness and generality of our method. As illustrated in Fig.~\ref{fig:Examples}, the proposed model successfully reconstructs two distinct irradiance patterns on Plane $A$ and Plane $B$, respectively, demonstrating consistent performance across different target configurations. The simulated results clearly show that the lens can accurately reproduce the prescribed images on both planes, confirming the effectiveness of the double-freeform optimization in handling diverse angular-spatial mappings.

It is worth noting that, in the last row of Fig.~\ref{fig:Examples}, the image on Plane $B$ appears to have a lower overall brightness compared with that on Plane $A$. This difference, however, does not imply any physical loss of optical power during propagation. Instead, it reflects the redistribution of the available light energy according to the normalized irradiance level defined by the target pattern on Plane $B$. The total flux is conserved; only the relative irradiance has been rescaled to match the specified target brightness distribution.

Fig.~\ref{fig:High_Contrast} provides a more challenging example, where the two images from the last row of Fig.~\ref{fig:Examples} are contrast-enhanced and used as the target irradiance distributions. Owing to the higher contrast and increased complexity of the targets, this case requires longer propagation distances than the previous example. As shown in Fig.~\ref{fig:High_Contrast}, when the overall optical system is more compact, the rays undergo stronger deflection and the optimized freeform surfaces exhibit more pronounced relief. In contrast, increasing the lens thickness and enlarging the separation between the two receptive planes significantly reduces the ray deflection, leading to smoother surface geometries. This example further demonstrates the effectiveness of our model for high-contrast and complex target patterns.

Our method can also be extended from collimated illumination to a point-source setting. For a Lambertian point source, the luminous flux is proportional to the solid angle rather than the projected area. Accordingly, Eq.~\eqref{eq:mass_consistency} is replaced by
\begin{equation}
E_{\text{area}}=\sum\nolimits_{\triangle \mathbf{v}_1^i \mathbf{v}_1^j \mathbf{v}_1^k\in\mathcal{F}_1}\left(S(\triangle \mathbf{v}_1^i \mathbf{v}_1^j \mathbf{v}_1^k)-S^0_{ijk}\right)^2.
\end{equation}
Here $S(\cdot)$ denotes the solid angle subtended by the triangle with respect to the point source, and $S^0_{ijk}$ is the corresponding initial solid angle of triangle $\triangle \mathbf{v}_1^i \mathbf{v}_1^j \mathbf{v}_1^k$. As shown in Fig.~\ref{fig:Point_Source}, the incident and exit surfaces are initialized as two spherical surfaces with radii $7\,\mathrm{cm}$ and $9.5\,\mathrm{cm}$, respectively, both centered at the point source, so that the initialization is consistent with the spherical wavefront emitted by the source. The outermost extent of the lens subtends an angle of $40^\circ$ with respect to the optical axis, thereby forming a quadrangular pyramid of rays. The imaging regions are defined by intersecting this pyramid with the planes $z=-10\,\mathrm{cm}$ and $z=-20\,\mathrm{cm}$. Under these modifications, the proposed framework can still realize two prescribed irradiance patterns from a point-source input, demonstrating its extensibility to more general source configurations.

Transforming a Gaussian beam into a uniform collimated beam is a fundamental topic in beam shaping. Our method can readily address this task in a straightforward manner. In the design stage, we set identical Gaussian irradiance distributions as the target patterns on both Planes $A$ and $B$, leading the optimization to naturally yield a self-consistent mapping that preserves collimation. As illustrated in Fig.~\ref{fig:Gaussian}, the optimized double-freeform lens first converts a uniform collimated beam into a Gaussian irradiance profile (top row), which remains stable over a certain propagation distance.
The same lens is then applied in reverse—without any modification or re-optimization—to transform a Gaussian input beam back into a uniform collimated output (bottom row).
This bidirectional experiment demonstrates the physical reversibility and versatility of the proposed design in realizing Gaussian–uniform beam conversion through a single optical element.

\section{Conclusion}
In conclusion, this study presents a unified double-freeform lens design framework that enables simultaneous angular and spatial control of light within a single refractive element.
By jointly optimizing the geometries of the incident and exit surfaces under dual-plane irradiance constraints, the proposed method precisely modulates light propagation without requiring complex multi-element systems.
Comprehensive simulations demonstrate accurate irradiance reproduction and physically consistent directional control across diverse target configurations, validating the effectiveness of the approach.
Beyond reducing system complexity, the framework establishes a versatile foundation for compact, high-performance optical systems in illumination engineering, beam shaping, and laser processing.
Future research will extend this methodology toward wavelength-dependent and polarization-sensitive designs, as well as machine-learning-assisted optimization, further advancing the integration of computational and physical optics. 

\section*{Funding}
This research was supported by the National Natural Science Foundation of China (No.62272433), Anhui Provincial Natural Science Foundation (No.2508085ZD011) and the Fundamental Research Funds for the Central Universities.

\section*{Disclosures}
The authors declare no conflicts of interest.

\section*{Data Availability}
Data underlying the results presented in this paper are not publicly available at this time but may be obtained from the authors upon reasonable request.

\bibliography{references}
\end{document}